\documentstyle[preprint,aps]{revtex}

\begin{document}
\draft
\title{Thermally assisted quantum cavitation in  solutions
of $^3$He in $^4$He}
\author{Dora M. Jezek$^1$, Montserrat Guilleumas$^2$,
Mart\'{\i} Pi$^3$, and Manuel Barranco$^3$.}
\address{$^1$Departamento de F\'{\i}sica, Facultad de Ciencias Exactas
y Naturales, \\
Universidad de Buenos Aires, RA-1428 Buenos Aires, Argentine}
\address{$^2$Dipartimento di Fisica, Universit\`a di Trento. 38050
Povo, Italy}
\address{$^3$Departament d'Estructura i Constituents de la Mat\`eria,
Facultat de F\'{\i}sica, \\
Universitat de Barcelona, E-08028 Barcelona, Spain}

\maketitle

\begin{abstract}

We have investigated the quantum-to-thermal crossover temperature
T$^*$ for cavitation in liquid  helium mixtures
up to $ 5 \%$   $^3$He concentrations. With respect to
the pure    $ ^4$He   case,    T$^* $ is sizeably reduced,
 to a value below  50 mK for
   $^3$He concentrations above $ 2 \% $. As in pure $^4$He,
the homogeneous cavitation  pressure is systematically found
close to the  spinodal pressure.

\end{abstract}


Quantum cavitation in superfluid liquid $^4$He has been experimentally
observed by Balibar and coworkers \cite{ba1}. They have used
 an ultrasound technique by means of which, a large
pressure oscillation is created in a small bulk region
of the experimental sample. This
prevents cavitation on the walls of the experimental
cell. Moreover, as superfluid
 $^4$He can be made perfectly pure, heterogeneous cavitation can be
avoided. Due to the
 nature of these experiments, it is very difficult to
 determine the  pressure (P) and temperature (T) at the cavitation site.
Consequently, a quantitative analysis of the experimental
results relies on
theoretical estimates of the equation of state and cavitation
   barriers in the spinodal region \cite{ma1,gui1}. Depending
  on these theoretical inputs and on experimental conditions,  such
as volume
  and observation time, it is inferred that quantum cavitation likely
 takes  over thermal cavitation at a
temperature in the  200-240 mK range. The agreement between experiment
\cite{ba1} and theory \cite{ma1,gui1} can be considered as
 satisfactory given the inherent limitations of both.

In this work we address the problem of thermally assisted quantum
cavitation in low $^3$He-concentration, liquid helium mixtures (up to
   $ 5 \% $). The relevance of this problem is twofold. First, the
planned experiments on quantum cavitation in these mixtures
 \cite{ba2}. Second, the complexity introduced
 in the theoretical description of the still superfluid liquid $^4$He,
  when a small but sizeable $^3$He amount is present.
 To our knowledge, no investigation of this
kind, either experimental or theoretical, exists in the literature.
Some effort has been concentrated in the study of supersaturated
$^3$He-$^4$He mixtures at positive pressures and low temperatures
(see for example Refs. \cite{lif1,sat1,bur1,jez1}), or near the
tricritical point (see Ref. \cite{hof1} and refs. therein).

The method we use here is conceptually simple and technically
workable. It
is based, on the one hand, in using a density functional \cite{da1} that
reproduces the thermodynamical
properties of the $^3$He-$^4$He liquid mixture at zero temperature.
This
functional has been slightly modified to reproduce the surface
properties of the $^3$He-$^4$He interface, and those of the mixture
free-surface \cite{gui2,gui3}. All these properties are relevant for a
quantitative description of the cavitation process.
On the other hand, we make use of a functional-integral approach
  especially well suited to find the crossover temperature
 T$^*$  with the only further approximation of imposing irrotational
motion for the growing of the critical bubbles.

Let us first generalize the functional-integral approach
 as used in \cite{gui1}, to the mixture
case (see also \cite{chu1}  for another example of how it applies).
We recall that above a crossover temperature T$^*$, the
cavitation rate, i.e., the number of bubbles formed per unit time and
volume, is given by
\begin{equation}
J_T = J_{0T}\, e^{-\Delta\Omega_{max}/T}\, ,
\label{eq1}
\end{equation}
where $\Delta\Omega_{max}$ is the barrier height for thermal activation.
 Below T$^*$, assuming $ \Delta \Omega_{max} >> $ T, one can write
\begin{equation}
J_Q = J_{0Q}\, e^{-S_{min}}\, ,
\label{eq2}
\end{equation}
being $S_{min}$  the minimum of the imaginary-time action
\begin{equation}
S(T) =  \oint {\rm d}\tau \int {\rm d}{\vec r}\,\, {\cal L}\, .
\label{eq3}
\end{equation}
${\cal L}$ is the imaginary-time classical Lagrangian density of
the system. For a given T,
the time-integration is extended over a "period"
$\tau$ in the
inverted barrier potential well
with $\tau=\hbar/$T.
The crossover temperature T$^*$ is determined from the angular
frequency $ \omega_p $ of the small amplitude
 oscillations around the minimum of the potential well (see
 \cite{chu2} and Refs. therein).
 The prefactors $J_{0T}$ and $J_{0Q}$ depend on the dynamics of
the cavitation process. In actual calculations, they are often
estimated as the number of cavitation sites per unit volume times
 an attempting frequency.

The Lagrangian density can be easily obtained from the
 zero-temperature density
functional of \cite{gui2,gui3}, whose use is justified in view of the
low temperatures involved (below 200 mK). The critical cavity
density profiles $\rho_3^0(r)$,  $\rho_4^0(r)$
are obtained solving the coupled Euler-Lagrange equations
\begin{equation}
\frac{\delta\omega(\rho_3,\rho_4)} {\delta\rho_q} = 0\,\,\,\,\, ,
q=3,4 \,\, ,
\label{eq4}
\end{equation}
where $\omega(\rho_3,\rho_4)$ is the grand potential density and
$\rho_q$ are the particle densities of each helium isotope.
$\Delta\Omega_{max}$ is given by
\begin{equation}
\Delta\Omega_{max} =\int {\rm d}{\vec
r}\left[\omega(\rho_3^0,\rho_4^0)-\omega(\rho_{m3},\rho_{m4})\right]\,
 ,
\label{eq5}
\end{equation}
where $\rho_{mq}$ are the corresponding densities of the
metastable homogeneous liquid (see \cite{gui2,gui3} for details).
Assuming that only spherical bubbles are developing, the collective
 velocities $ \vec u_q(\vec r, t) $ of both helium fluids are
irrotational and one can then define for each isotope a velocity
potencial field
 s$_q({\vec r},t)$ such that $\vec u_q(\vec r,t)$ $ \equiv $
$ \nabla $ s$_q({\vec r},t)$. It follows that

\begin{equation}
{\cal L} = \sum_q m_q \dot{\rho}_q s_q - {\cal H}(\rho_3,\rho_4,
s_3,s_4)\,\, ,
\label{eq6}
\end{equation}
where ${\cal H}(\rho_q,s_q)$ is the imaginary-time Hamiltonian
density
\begin{equation}
 {\cal H}(\rho_3,\rho_4, s_3,s_4)
 =\frac{1}{2} \sum_q m_q\rho_q{\vec u_q}\,^2
-\left[\omega(\rho_3,\rho_4)-\omega(\rho_{m3},\rho_{m4})\right]\,\, .
\label{eq7}
\end{equation}
Hamilton's equations yield  the following four equations :
\begin{equation}
m_q\dot{\rho_q} = \frac{\delta {\cal H}} {\delta s_q} = -m_q
\nabla(\rho_q {\vec u_q})
\label{eq8}
\end{equation}
\begin{equation}
m_q\dot{s_q} = -\frac{\delta{\cal H}} {\delta\rho_q}\, .
\label{eq9}
\end{equation}
Eqs. (\ref{eq8}) are the continuity equations. Taking the gradient of
Eqs.
(\ref{eq9}) one gets
\begin{equation}
m_q\frac{{\rm d}{\vec u_q}}{{\rm dt}} = -\nabla
\left\{
\frac{1}{2}m_q{\vec u_q}\,^2-\frac{\delta\omega}{\delta\rho_q}
\right\}\, .
\label{eq10}
\end{equation}

To determine T$^*$ one has to find
the small amplitude, periodic solutions of Eqs. (\ref{eq8}) and (\ref
{eq10}) linearized  around $ \rho_3^0 $ and $ \rho_4^0 $
\cite{gui1,chu1}. Defining the transition densities $ \rho_q^1(r) $ as
\begin{equation}
\rho_q(\vec r,t) \equiv \rho_q^0(r) + \rho_q^1(r)\, cos(\omega_p t)\,
\, ,
\label{eq11}
\end{equation}
where $\rho_q^1(r)$ is much smaller than $\rho_q^0(r)$, and keeping
only first order terms in ${\vec u_q}(r,t)$ and in $\rho_q^1(r)$, one
gets:
\begin{equation}
\omega^2_p \rho_q^1(r)= \frac{1}{m_q}  \nabla\left[\rho_q^0(r)
\nabla\left( \sum_{q'=3,4}
\frac{\delta^2\omega}{\delta\rho_q
\delta\rho_{q'}}\bullet\rho_{q'}^1(r)\right)\right]\,\,\,,
q=3,4.
\label{eq12}
\end{equation}
In Eq. (\ref{eq12}), \,\,  $\frac{\delta^2\omega}{\delta\rho_q
\delta\rho_{q'}}\bullet\rho_{q'}^1(r)$ means that
 $\delta\omega / \delta\rho_q  $
 has to be linearized, keeping only terms in
 $\rho_{3}^1 $ and $\rho_{4}^1$, and their derivatives.

   Eq. (\ref{eq12}) is a fourth-order linear differential, eigenvalue
equation for the "vector" $ (\rho_3^1(r), \rho_4^1(r)) $, whose
right-hand side term is straigthforward but very cumbersome to obtain.
We have done it using the Mathematica software  package \cite{math}.
 As in \cite{gui1}, a careful analysis shows that the
  physical solutions of Eqs. (\ref{eq12}) have to fulfill
$(\rho_q^1)'(0)=(\rho_q^1)'''(0)=0$, and have to fall exponentially
to zero at large distances. From the linearized continuity equation
$\rho_q^1(r)\propto -\nabla(\rho_q^0 {\vec u_q})$,
it is obvious that the integral of $ \rho_q^1(r) $ over the whole
space is zero.

  We have solved the eigenvalue  Eq. (\ref{eq12}) as in \cite{gui1}.
For a given pressure and $^3$He-concentration
$ x \equiv  \rho_{m3}/(\rho_{m3} + \rho_{m4}) $, only a
positive eigenvalue $ \omega^2_p $ is found, out of which we get  T$^*
=\hbar \omega_p / 2 \pi $.

Fig. 1 shows T$^*$
(mK) as a function of P(bar) for $x=$ 0.1, 1, 2, 3, 4 and 5 $ \% $.
Compared to the pure $^4$He case \cite{gui1},
T$^*$(P) has now a more complex structure.
It is worth to note that the maximum of the T$^*$(P) curve has decreased
from $ \sim $240 mK for pure $^4$He \cite{gui1} down to $\sim $ 140 mK
 for $^3$He-concentrations as small as $ 1 \%$.

Fig. 2 shows two different bubble configurations for x$=1 \% $.
Configuration (a) corresponds to P$=-8.23$ bar and T$^* = 62.7$ mK, and
configuration (b) to P$=-5.17$ bar and T$^* = 115.7$ mK.
  The solid lines represent the $^3$He and $^4$He critical bubble
  particle densities in $ {\rm \AA}^{-3}$, and the  dashed
(dash-dotted) lines represent
$\rho_3^1(r) \,\, (\rho_4^1(r)) $ in arbitrary units. Near the
spinodal region, the "bubble" configuration is filled with $^3$He: the
surface tension that matters for bubble formation is that of the
$^3$He-$^4$He interface.
Away from the spinodal region (configuration (b)), the critical
bubble is a true bubble covered with $^3$He
(Andreev states): the surface tension that matters now is that of
 the $^3$He-$^4$He liquid free-surface, which is about ten
times larger than the previous one. The different surface tensions
involved in these processes, together with the existence
of a $^3$He-$^4$He saturation curve at negative pressures down to
$ x \sim  2.4 \%$ \cite{gui2} are the cause of the structures
displayed in Fig. 1.
 We will not give here any further detail
since, as in the pure $^4$He case, only the part of the
  T$^*$(P) curve near the spinodal
region is relevant for the cavitation problem.

 It is interesting to see that the
transition densities $ \rho_q^1$ evolve from those corresponding to
"volume oscillations" (Fig. 2a) to "surface oscillations"
for $^4$He, and a mixed surface-volume type for $^3$He
(Fig. 2b), to eventually become pure "surface oscillations"
for both isotopes  when moving from
the spinodal towards the saturation line. This fact has been
already found in pure $^4$He \cite{gui1}.

To establish which T$^*$(P) in Fig. 1 corresponds to that of the
experimental conditions, we have to determine the homogeneous
cavitation pressure P$_h$, which is the one
the system can sustain before bubbles nucleate at an appreciable
rate.  P$_h$ can be obtained  solving the equation
\begin{equation}
1 = (Vt)_e\, J \, ,
\label{eq13}
\end{equation}

\noindent taking J = J$_T$ and
J$_{0T} = (k_B T)/(h V_0) $ if T$>$T$^*$, where $ k_B T / h $ is
a thermal attempting frequency, and
$V_0 = 4 \pi R^3_c/3$ is the inverse of the number of possible
nucleation centers per unit volume, with  R$_c$=10
$\AA$ as a typical radius of the critical bubble. For T
$\leq$ T$^*$, one takes J $=$ J$_{Q}$ and, lacking of a better choice,
  J$_{0Q} =$ J$_{0T}$(T$=$T$^*$). It yields J$_{0Q}$ of  the
order of  10$^5 {\rm \AA}^{-3}$ s$^{-1}$.
 Another possible guess for the quantum attempting frequency is to
equal it to $ v_s / R_c $, where $v_s$ is the sound velocity,
which is of the
order of $ 10^{12} {\rm \AA} \,$ s$^{-1} $ in the region of interest.
This choice differs from the previous one in two orders of magnitude.
However, this difference in the prefactor value
 does not alter in practice the results. Indeed, we have chosen two
 extreme experimental values for the  factor
(Vt)$_e$ (experimental volume $\times$ time) which differ ten orders
of magnitude without much influencing the obtained T$^*$-value.

 P$_h$(T$^*$) is shown as circles (squares) on the
 curves in Fig 1. The circles correspond to
(Vt)$_e =$ 10$^4 {\rm \AA}^3 $ s,  and the squares to
 10$^{14} {\rm \AA}^3 $ s. Compared to the pure $^4$He case
\cite{gui1}, and depending on the (Vt)$_e$-value,
for $x = 1 \% $ T$^*$ has been reduced a factor of 4 or 5,
respectively.

Fig. 3  shows P$_h$ as a function of T for the above
mentioned $x$-values and
(Vt)$_e =$ 10$^{14} {\rm \AA}^3 $ s. Thermal and quantum regimes are
displayed. The dashed line is  the extrapolation of the
thermal regime
to temperatures close to T$=$0. Finally, Fig. 4 displayes the
T$=$0, P$_{sp}$(x) spinodal line together with the P$_h$(x,T$^*$)
curves for the indicated (Vt)$_e$ values.
Notice that the smallest $x$-value displayed in this figure is
$0.1 \%$. For $^3$He concentrations closer to zero, it is very
unlike that $^3$He has time enough to difuse and develop the critical
configurations that constitute the starting point of the present
calculations. Also shown in that figure (dots on the y-axis) are the
pure $^4$He values \cite{gui1}.

In conclusion, we have thoroughly investigated
thermally-assisted quantum cavitation in $^3$He-$^4$He liquid
mixtures at small $^3$He concentrations. Our quantitative  predictions
rely on a robust density functional that reproduces the relevant
characteristics of the mixture and its interfaces, and on a sound
formalism, the functional integral theory. Besides the approximations
inherent to the method (use of a density functional,
 irrotational flows), no further
approximation has been introduced to obtain the quantum-to-thermal
crossover temperature. The present results might thus be a
valuable guide to the planned experiments \cite{ba2}.

We would like to thank Sebastien Balibar, Eugene Chudnovsky and Jacques
Treiner for useful discussions. This work has been supported in part by
the CICYT, by the IN2P3-CICYT exchange program, by the
 Generalitat de Catalunya Visiting Professors  and
GRQ94-1022 programs, and by the CONICET (Argentine) Grant No. PID 97/93.

\begin{figure}
\caption{  T$^*$ as a function of P.
The homogeneous cavitation pressure
 P$_h$(T$^*$) is shown as circles (squares) for
(Vt)$_e =$ 10$^4 {\rm \AA}^3 $ s
 (10$^{14} {\rm \AA}^3 $ s), respectively.}

\label{fig1}
\end{figure}
\begin{figure}
\caption{
  Particle densities
$\rho_4^0(r)$ and $\rho_3^0(r)$ of the critical bubbles
(solid lines), and the
$\rho_4^1(r)$ (dash-dotted lines) and $\rho_3^1 (r)$ (dashed lines)
transition densities for x$=1 \%$, corresponding to:
 (a)  P$=-8.23$ bar and T$^* = 62.7$ mK.
(b) P$=-5.17$ bar and T$^* = 115.7$ mK.
$\rho_q^1(r)$ are drawn in arbitrary units, $\rho_q^0(r)$ in
\AA$^{-3}$. }
\label{fig2}
\end{figure}
\begin{figure}
\caption{ Homogeneous cavitation pressure
 P$_h$ as a function of T for the same
 x-values as in Fig. 1.}
\label{fig3}
\end{figure}
\begin{figure}
\caption{
 The  T$=$0 spinodal line P$_{sp}$ (x) (solid line), and
the homogeneous cavitation pressure P$_h$(x,T$^*$)
 (dashed and dash-dotted curves) for the indicated (Vt)$_e$ values.
The dots on the y-axis are the corresponding pure $^4$He values.}
\label{fig4}
\end{figure}

\begin{references}
\bibitem{ba1}S. Balibar, C. Guthmann, H. Lambare, P. Roche, E.
Rolley and H.J. Maris, J. Low Temp. Phys. {\bf 101}, 271 (1995).
\bibitem{ma1} H.J. Maris, J. Low Temp. Phys. {\bf 98}, 403
(1995).
\bibitem{gui1}M. Guilleumas, M. Barranco, D. Jezek, R. Lombard and
M. Pi, to be published in Phys. Rev. B (1996).
\bibitem{ba2} S. Balibar, private communication.
\bibitem{lif1} I.M. Lifshitz, V.M. Polesskii and V.A. Khokhlov,
Sov. Phys. JETP {\bf 47}, 137 (1978).
\bibitem{sat1} T. Satoh, M. Morishita, M. Ogata and S. Katoh,
 Phys. Rev. Lett. {\bf 69}, 335 (1992).
\bibitem{bur1} S.N. Burmistrov, L.B. Dubovskii and V.L. Tsymbalenko,
J. Low Temp. Phys. {\bf 90}, 363 (1993).
\bibitem{jez1} D.M. Jezek, M. Guilleumas, M. Pi and
M. Barranco, Phys. Rev. {\bf B51}, 11981 (1995).
\bibitem{hof1} J.K. Hoffer and D.N. Sinha, Phys. Rev. {\bf A33},
1918 (1986).
\bibitem{da1} F. Dalfovo and S. Stringari, Phys. Lett {\bf 112A}, 171
(1985).
\bibitem{gui2}M. Guilleumas, D. Jezek, M. Pi,
M. Barranco and J. Navarro, Phys. Rev. {\bf B51}, 1140 (1995).
\bibitem{gui3}M. Guilleumas,  M. Pi,
M. Barranco, D. Jezek and J. Navarro, Phys. Rev. {\bf B52}, 1210
(1995).
\bibitem{chu1} E. M. Chudnovsky, A. Ferrera and A. Vilenkin, Phys.
Rev.
{\bf B51}, 1181 (1995).
\bibitem{chu2} E. M. Chudnovsky, Phys. Rev. {\bf A46}, 8011 (1992)
\bibitem{math} S. Wolfram, {\it Mathematica, A System for Doing
Mathematics by Computer }, (Addison-Westley Publishing Company, Inc.,
California, 1991).
\end{references}
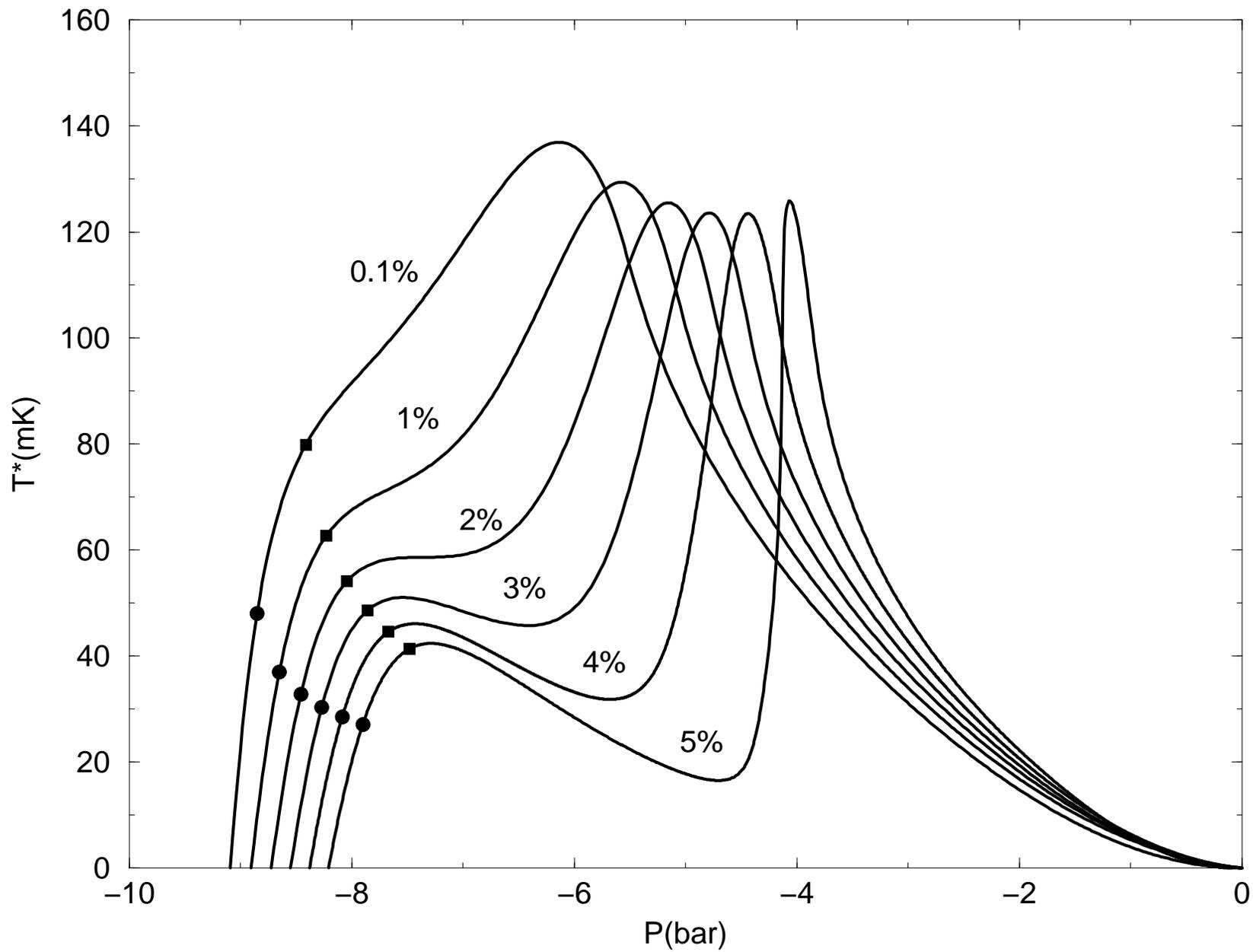
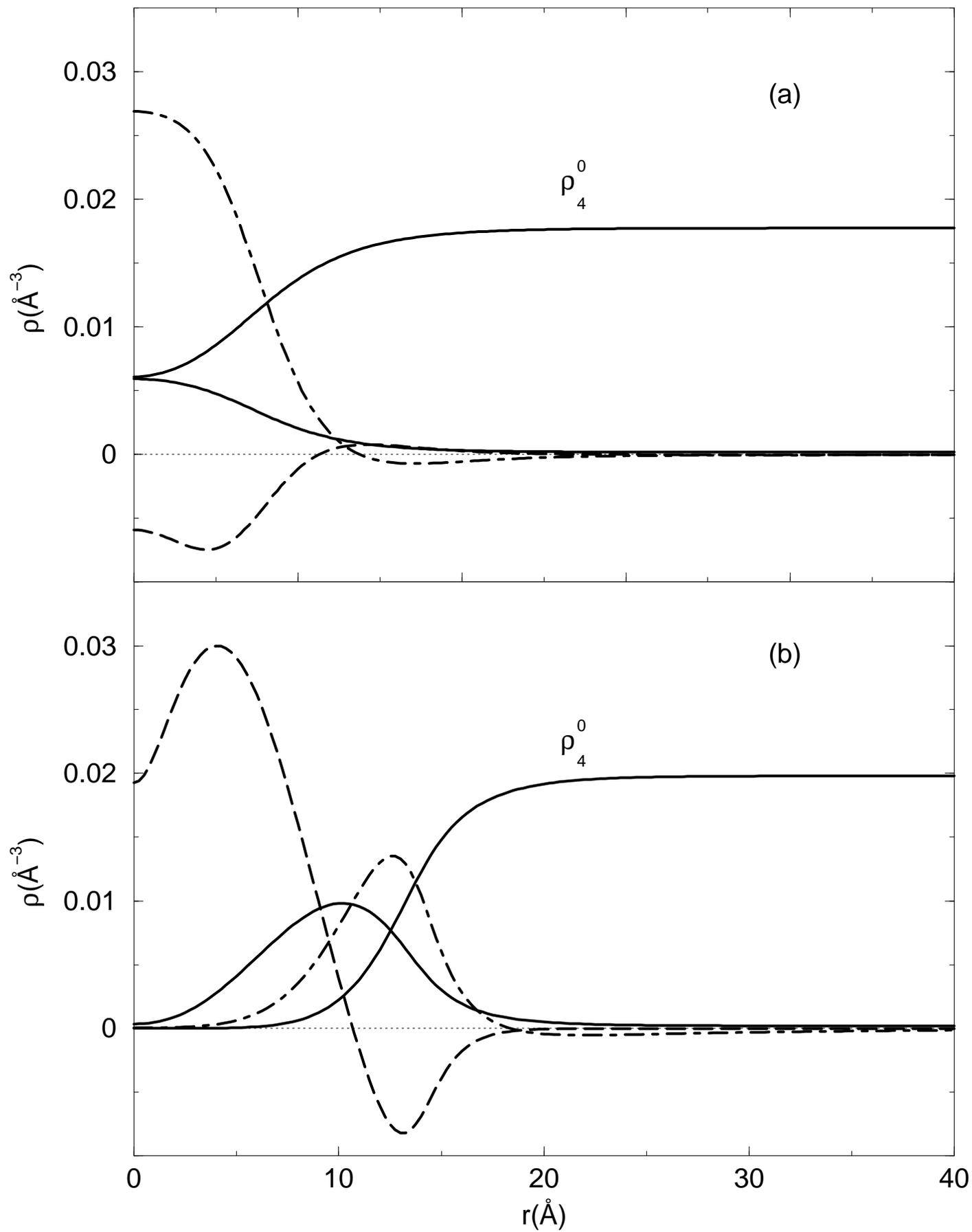
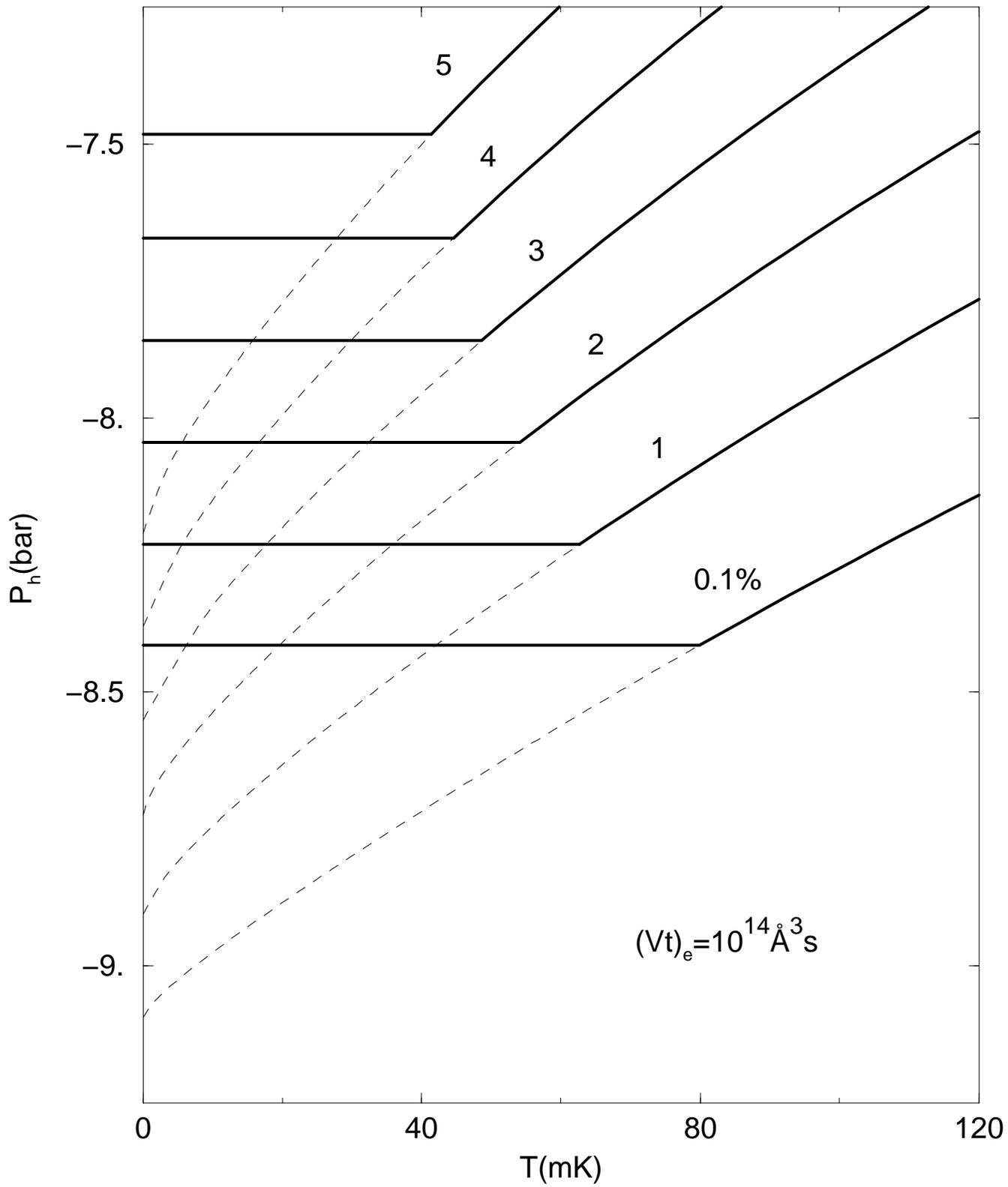
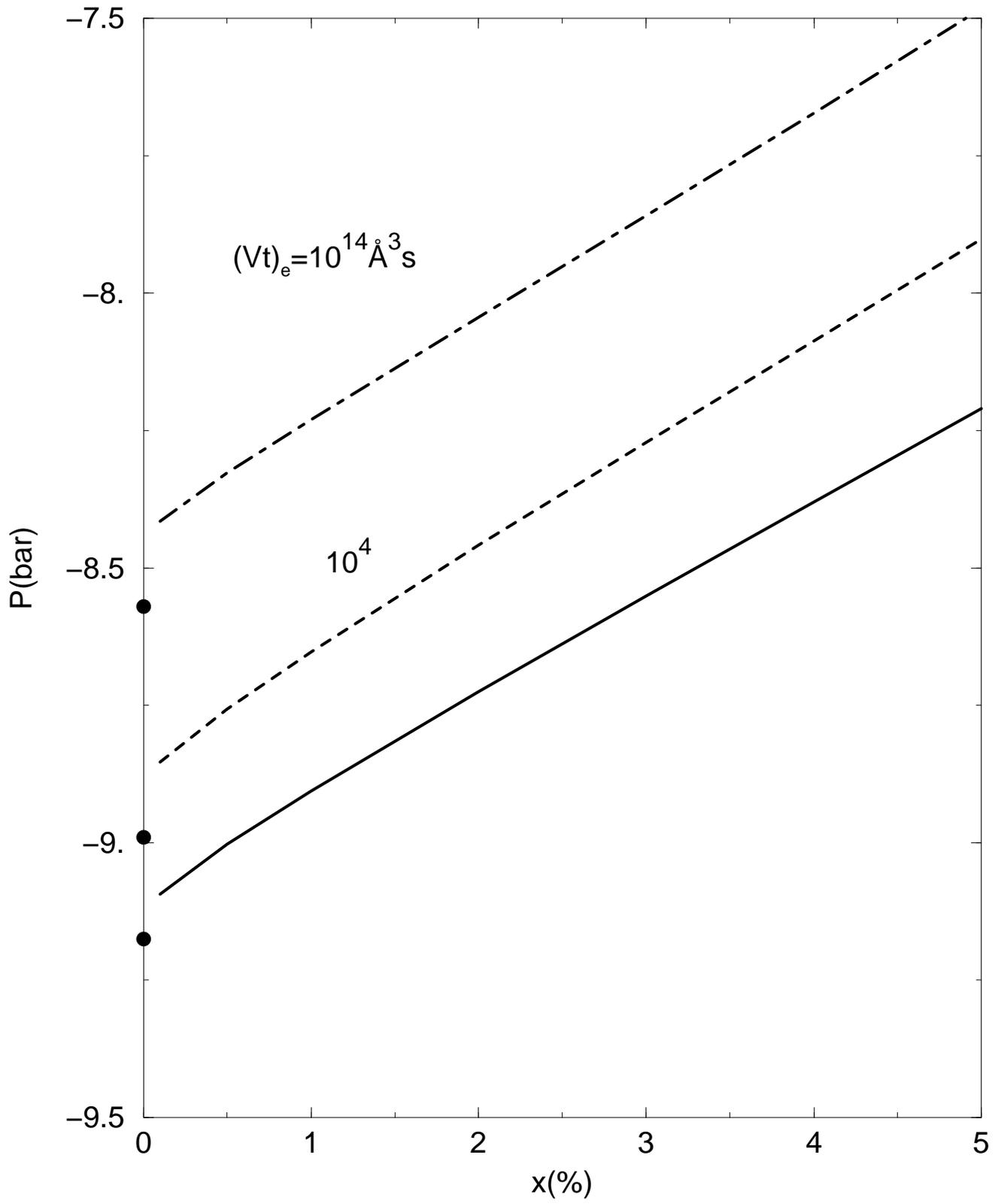
\end{document}